# Rapidly fluctuating orbital occupancy above the orbital ordering transition in spin-gap compounds

B. Rivas-Murias,[1,2] H. D. Zhou,[3] J. Rivas,[4] F. Rivadulla[1,2,*]

*[1]Department of Physical Chemistry and [2]Center for Research in Biological Chemistry and Molecular Materials, University of Santiago de Compostela, 15782 Santiago de Compostela, Spain*

*[3]National High Magnetic Field Laboratory, Florida State University, Tallahassee, FL 32306-4005, USA*

*[4]Department of Applied Physics, University of Santiago de Compostela, 15782 Santiago de Compostela, Spain*

***e**-mail: f.rivadulla@usc.es*

Several spin systems with low dimensionality develop a spin-dimer phase within a molecular orbital below $T_S$, competing with long-range antiferromagnetic order. Very often, preferential orbital occupancy and ordering are the actual driving force for dimerization, as in the so-called orbitally-driven spin-Peierls compounds ($MgTi_2O_4$, $CuIr_2S_4$, $La_4Ru_2O_{10}$, $NaTiSi_2O_6$, etc.). Through a microscopic analysis of the thermal conductivity κ (T) in $La_4Ru_2O_{10}$, we show that the orbital occupancy fluctuates rapidly above $T_S$, resulting in an orbital-liquid state. The strong orbital-lattice coupling introduces dynamic bond-length fluctuations that scatter the phonons to produce a κ (T)~T (*i.e.* glass-like) above $T_S$. This phonon-glass to phonon-crystal transition is shown to occur in other spin-dimer systems, like $NaTiSi_2O_6$, pointing to a general phenomenon.

Preferential orbital occupation in transition-metal compounds shapes the chemical bond of these systems, and is the origin of many different long-range magnetic and charge arrangements.[1] In magnetically frustrated structures, or slightly correlated low dimensional systems with small spin (S<1), long-range ordering of spin-dimers (singlets) competes with antiferromagnetic (AF) order,[2] as in pyroxene $NaTiSi_2O_6$ (S=1/2)[3]. In layered $La_4Ru_2O_{10}$ (S=1), Wu *et al.*[4] proposed that an orbital ordering: $[dxz^{\uparrow\downarrow}, dxy^{\uparrow}, dyz^{\uparrow}]$ introduces a strong exchange anisotropy that drives the system electronically 1D, and forms $Ru^{4+}$-$Ru^{4+}$ spin-singlets below $T_S \approx 160$ K (a realization of an orbitally driven Peierls state).[5] This anisotropy is reflected in alternating short/long Ru-O bond-lengths in the low temperature (spin-gapped) phase (see Figure 1). However, these bond-distances are also compatible with an S≈0 at each $Ru^{4+}$ site $(dxz^{\uparrow\downarrow}, dxy^{\uparrow\downarrow}, dyz^{0})$, as originally proposed by Khalifah *et al.*[6]; and to distinguish between these scenarios is a difficult task.[7]

At high temperature, $T > T_s$, the undistorted Ru-O octahedra observed by neutron diffraction have been assigned to an equal orbital occupation that quenches the orbital degrees of freedom. But these equal bond-lengths could also be a consequence of a rapid fluctuation in the orbital occupancy, which does not quench the orbital channel but results in an average Ru-O bond. This subtle difference between equal vs. rapidly fluctuating orbital occupancy is really important, as it should change the way we describe the bond in solids above a static long-range orbital-order transition. For example, fluctuations in the orbital occupancy should manifest as variations in the bond-length that must influence the propagation of phonons, and according to the Goodenough-Kanamori rules[1] should introduce a rapid fluctuation in the sign of the magnetic exchange.

Here we report the analysis of thermal conductivity in $La_4Ru_2O_{10}$ and $NaTiSi_2O_6$, and show that their high-temperature phase above $T_S$ is better described by a mixture or a resonance between different structures determined by rapid fluctuating orbital occupancies, instead of an average occupancy that quenches the orbital degrees of freedom. This effect leads to a dramatic change in the temperature dependence of κ (T), that shows a phonon-crystal to phonon-glass transition at $T_S$.

The crystal structure of $La_4Ru_2O_{10}$ can be described as zigzag chains of corner sharing $RuO_6$ along *b* axes in the *ab* plane (see Figure 1).[6] The high-temperature (HT) monoclinic phase is characterized by equal Ru-Ru distances; in the low-temperature (LT) triclinic phase there is a lengthening (shortening) of the bond distance along the x (y) direction (Figure 1).[6] The orbital manifold of low spin $Ru^{4+}$ ($4d^4$) splits in an octahedral environment into a lower energy $xz^{\uparrow\downarrow}$ and a degenerate $xy^{\uparrow}$ $yz^{\uparrow}$. An ab-initio calculation proposed that the energy difference between the doubly occupied orbital and the single occupied doublet increases from ≈50 meV in the HT phase to ≈300 meV in the LT phase.[4] The strong exchange anisotropy induced by the orbital ordering forms a Ru-Ru molecular-orbital along the short bonds, opening a spin gap in the low-temperature phase.[4] On the basis of this interpretation $La_4Ru_2O_{10}$ has been proposed as a rare example of a spin-dimer system with S>1/2 and 2D structure, due to an unusually strong orbital anisotropy in a 4d system. Optical conductivity data by Moon *et al.*[8] supported this interpretation, over the initial proposal of $xy^{\uparrow\downarrow}$ $yz^{\uparrow\downarrow}$ (S=0 at $Ru^{4+}$ sites) in Ref. [6].

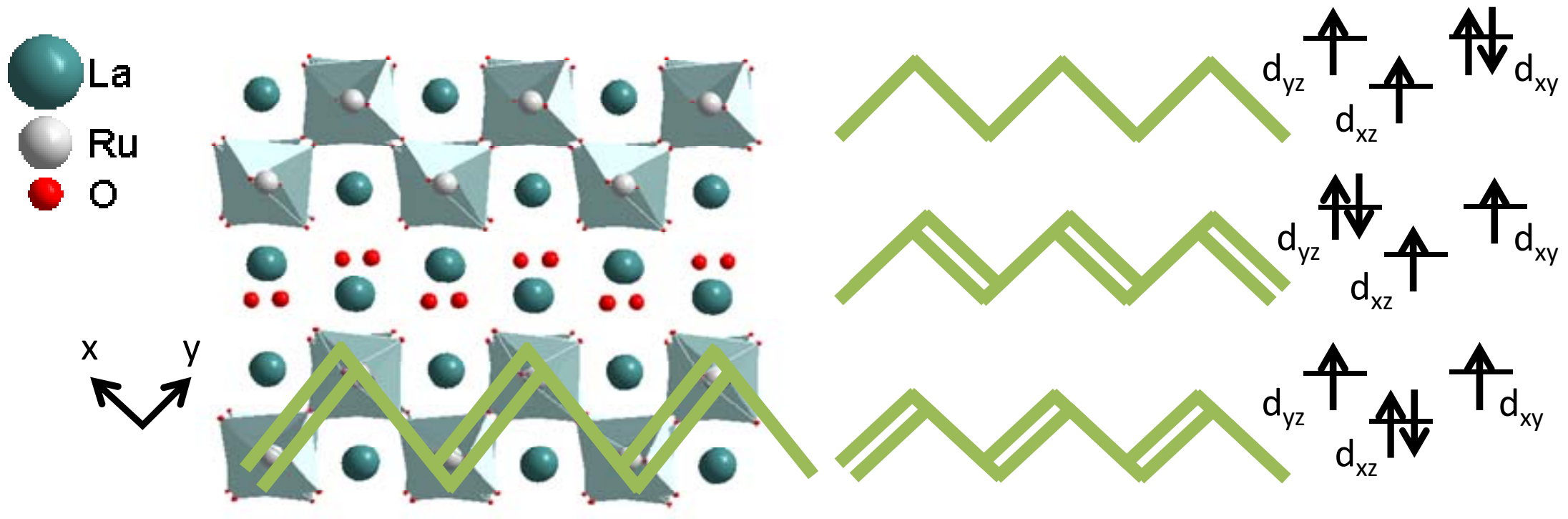

**Figure 1.** Left: Crystal structure of $La_4Ru_2O_{10}$ (*ab* plane viewed along the *c* direction). The solid lines show the short (double) and long (single) bonds in the low temperature monoclinic phase. There is another single bond along the c-axis (not shown). Right: Spin-orbital configurations of the $t_{2g}$ orbital set in $Ru^{4+}$, along with the resulting bond-configuration in the xy plane. Double (single) bonds represent short (long) Ru-O-Ru bonds. For the orbital configuration on the top there is another short bond along the c-direction (not shown).

In Figure 2, we show the temperature dependence of the magnetic susceptibility for $La_4Ru_2O_{10}$ at different pressures. The transition temperature increases with pressure at a rate $\partial \ln T/\partial P = 3.83 \times 10^{-2}$ GPa$^{-1}$. This increase with pressure is consistent with an interionic exchange coupling of the form $J \sim t^2/U$, hence supporting the formation of spin dimers instead of the $Ru^{4+}$ S=0 state. Also, although we did not measured experimentally the bulk modulus, B, for $La_4Ru_2O_{10}$, it must be similar to $Ba_4Ru_3O_{10}$ (B=113 GPa$^{-1}$),[9] given their similar coordination. Assuming this value, the Bloch's parameter[10] $\alpha = \partial \ln T_s/\partial V \approx 4.2$. The calculated α for AF insulators with direct *d-d* overlap is 3.3, and 4.6 for a *d-p-d* transfer.[11] This volume dependence of the susceptibility shows that the most probable path for the exchange interaction goes through the O:2p orbitals, that is, the spin-singlet state involves the Ru-O-Ru atomic trimer.

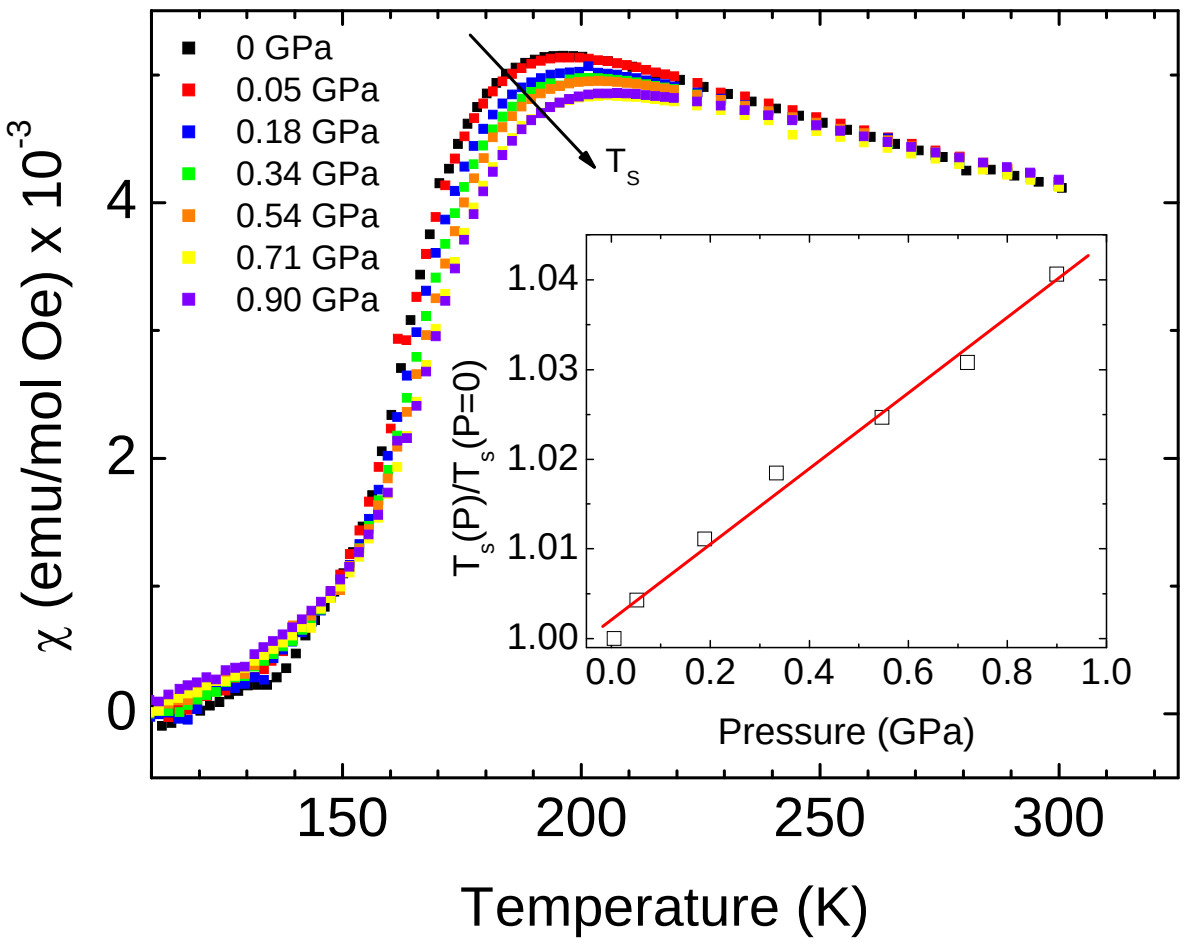

**Figure 2.** Temperature dependence of the magnetic susceptibility for $La_4Ru_2O_{10}$ at different pressures. Inset: Evolution of the transition temperature $T_S$ with pressure.

On the other hand, the assumption that orbital degrees of freedom are quenched above $T_s$ left the HT phase practically unexplored. However, with such a small energy difference ≈50 meV between the xz $^{\uparrow\downarrow}$ and xy$^{\uparrow}$ yz$^{\uparrow}$ states in the HT phase, fast fluctuations in the orbital occupancy are highly probable. Actually, Moon *et al.*[8] observed a gradual development of the in-plane anisotropy above $T_S$, and suggested that short-range orbital or spin correlations could survive above the spin gap temperature.

Due to the anisotropy in the bond-distance induced by the orbital ordering (see Figure 1, right), a rapid fluctuation in the orbital occupancy should manifest in a fluctuation in the bond-distance, undetectable in a conventional diffraction experiment.

However, other magnitudes like thermal conductivity or the speed of sound should be sensitive to these bond-length fluctuations, if present. The temperature dependence of the thermal conductivity for $La_4Ru_2O_{10}$ is shown in Figure 3. There is a dramatic change in the mechanism of heat-conduction at the magneto-structural transition at $T_S$. Given the large electronic resistivity, the cause of this effect must be in a fundamental change of the lattice contribution to the thermal conductivity. The behavior of the LT-phase is typical for a crystalline material, with a phonon peak at the maximum phonon mean-free path, Λ, and the characteristic reduction of $\kappa(T>T_{max})$ as temperature increases due to Umklapp scattering.[12] However, κ changes drastically its temperature dependence above $T_s$, increasing linearly with T in the HT phase as it is expected for an amorphous solid.[12] Although the absolute values of κ do not change appreciably, it is clear that spectra of low energy excitations of the lattice change completely in the HT phase with respect to the LT spin-dimerized state.

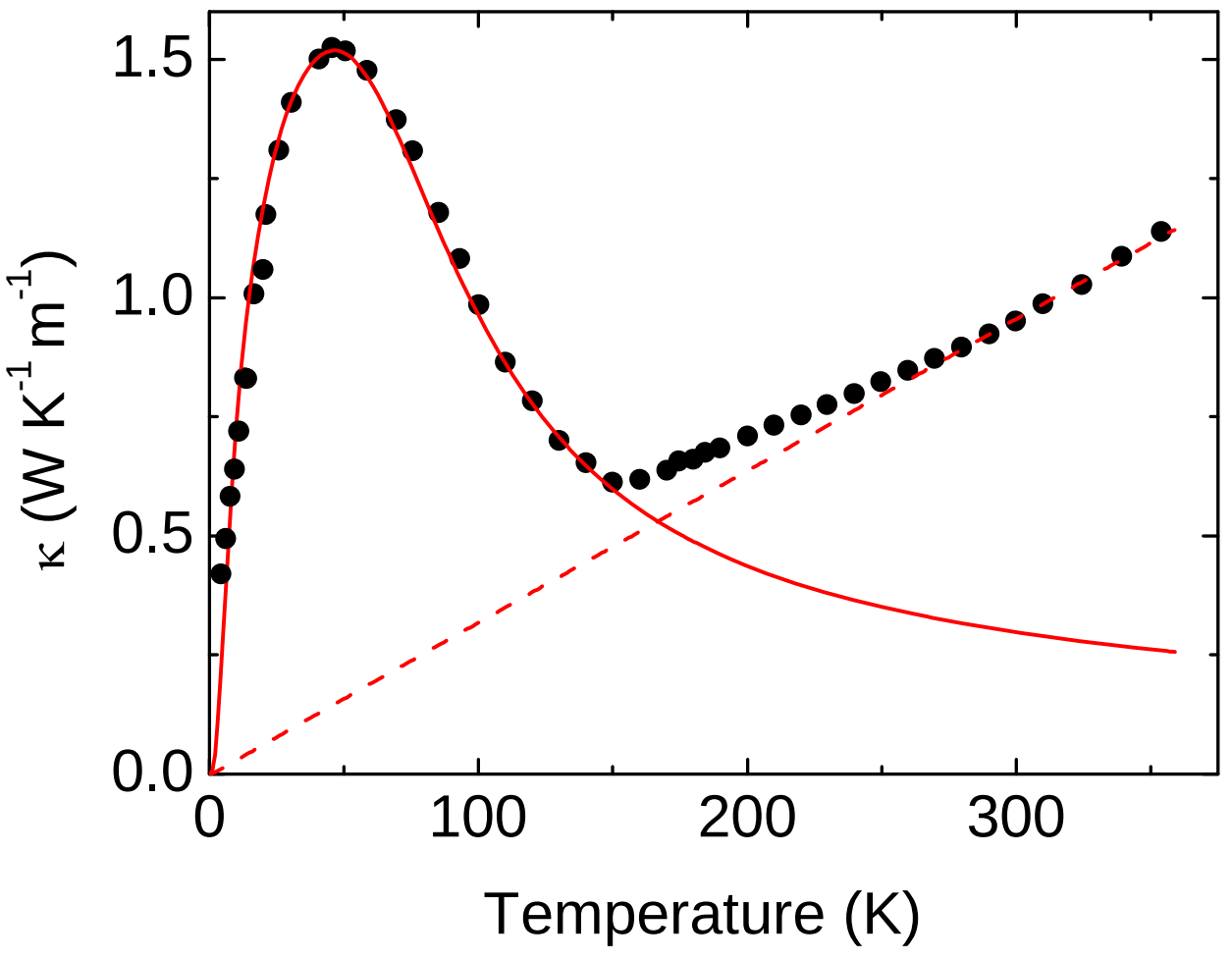

**Figure 3.** Temperature dependence of the thermal conductivity in $La_4Ru_2O_{10}$. Solid and dashed lines represent the fittings to equation (1) in the low and high temperature regimes, respectively, as explained in the text.

In a first approximation, the thermal conductivity can be estimated by the Debye model:[13]

$$\kappa(T) = \frac{1}{2\pi^2 v}\int_0^{\omega_D} \tau(\omega)\frac{\hbar\omega^4}{k_B T^2}\frac{e^{\hbar\omega/k_BT}}{\left(e^{\hbar\omega/k_BT}-1\right)^2}d\omega \qquad (1)$$

where $v$ and $\omega_D$ are the sound velocity and the Debye frequency, respectively, and the other symbols have their usual meaning. The relaxation time, τ(ω) is the sum of the contribution of different processes that produce an effective phonon scattering. In our analysis we have included three different terms:

$$\begin{aligned} &\tau^{-1}(\omega) = A + B\omega^2 T e^{-\Theta/\alpha T} + C\omega^n \\ &A = \frac{v}{1.8d} \\ &B = \frac{\hbar\gamma^2}{Mv^2\Theta} \end{aligned} \qquad (2)$$

The first term represents the scattering by grain boundaries in polycrystalline materials (d ≈1±0.2μm is the grain size obtained from electron microscopy, while the average phonon velocity, $v$, was stimated from the Debye temperature, Θ); the second term represents the Umklapp processes[14] (M and γ are the average atomic mass and the Grïneissen constant, that we assumed ≈2, respectively), and the last one is due to scattering by defects on the structure. For lattice-point defects, or any other object with a typical size smaller than the phonon wavelength, this reduces to Rayleigh scattering, having n=4 and

$$C = \frac{V_0}{4\pi v^3} \sum_i x_i S_i^2 \quad (3)$$

where $V_0$ and $x_i$ are the volume and molar fraction of defects. *S* represents the contributions of the local variation of mass, bond-distance and energy due to the defects on the host lattice.[15] Domain boundaries in the regular propagation of short-long-short-long Ru-O bonds in the LT triclinic structure (see Figure 1, right), will create scattering centers of ≈ 8 Å in linear dimensions, and a 5% variation in the local bond-distance with respect to the regular structure.

The fitting of experimental κ(T) to equation (1) is shown in Figure 3 (solid line). For boundary and Umklapp scattering we used the parameters calculated from our experimental data ($A = 3.6\times10^9$ $s^{-1}$, $B = 9.9\times10^{-17}$ s $K^{-1}$ and used α=1 as a fitting factor), except for Θ ≈ 225 K, which is ≈30% larger that obtained directly from our specific heat, although exactly the same as obtained from *C*(T) by Malik *et al.*[16] Assuming a 4% of defects ($C = 4.2\times10^{-41}$ $s^3$) results in a remarkably good agreement between the predictions from the Debye model and the experimental data in the LT dimerized phase.

Above $T_S$, κ(T) departs completely from the expected behavior for a crystalline solid. Kittel[17] gave the first interpretation to this behavior, assuming a constant Λ for every phonon frequency in the glass due the geometrical disordered structure. This should be a reasonable approximation, at least for small phonon wavelengths. From an analogy with the kinetic theory of gases, the thermal conductivity of a crystalline solid can be expressed as:

$$\kappa = \frac{1}{3} C v \Lambda \quad (4)$$

where *C* is the lattice heat capacity. To obtain the lattice specific heat we have fitted the background of the experimental *C*(T) curve to a polynomial function. The magnetic contribution that results from the subtraction of this polynomial to the total heat capacity is shown in the inset to Fig. 4. Integration of this peak gives a magnetic entropy 8.56 J/mol K, only ≈6% smaller than the theoretical value for low spin $Ru^{4+}$: R ln(2S+1) = 9.13 J/mol K. This supports the validity of our estimation for the lattice heat capacity.

Therefore, from the experimental values of κ(T) and *C*(T), and using $v$=5300 $ms^{-1}$ obtained from the experimental Θ, and the theoretical density from the Rietveld fitting of the x-ray powder patterns, we have estimated the temperature dependence of Λ shown in Figure 4. Above $T_S$, Λ ≈ 20 Å, and remains practically constant up to room temperature. This should not be taken as an accurate absolute value of Λ, but as an order of magnitude, which in this case is roughly of the size of the Ru-O-Ru bonds. This result points to anisotropic Ru-O bond-length fluctuations as the most probably source of geometrical disorder that produces the glass-like $\kappa(T>T_S)$.

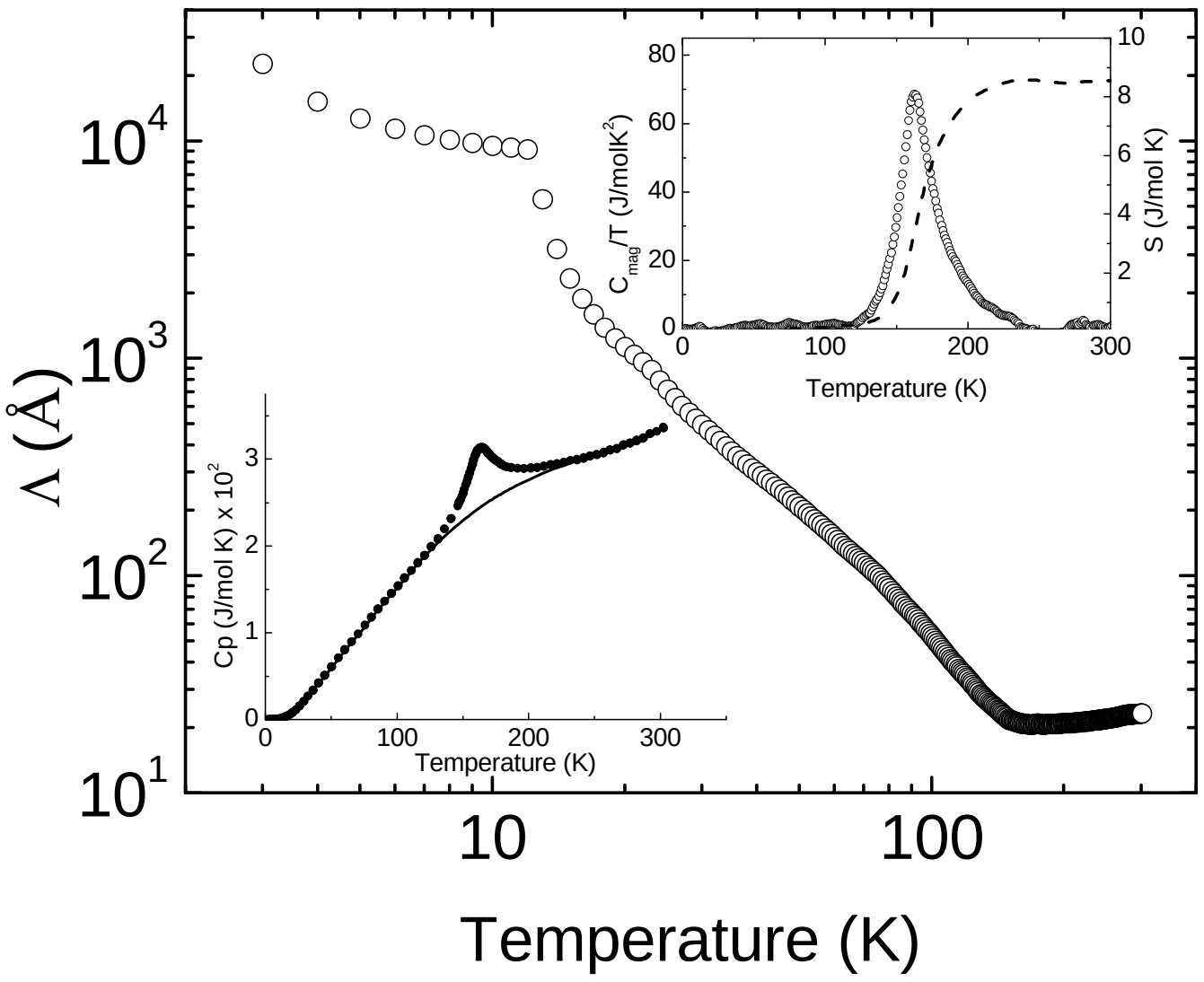

**Figure 4.** Temperature dependence of phonon mean free path in $La_4Ru_2O_{10}$. Lower inset: Total specific heat (circles) and estimated lattice background (line). Subtraction of the background to the total specific heat gives the magnetic contribution and the magnetic entropy (upper inset).

On the other hand, introducing 2.5% of holes into the Ru sites via $Ce^{+4}$-doping (Fig. 5) suppresses $T_S$ by 15 K and $\kappa(T_{max})$ drops ≈50%. This demonstrates the relevance of cooperative lattice distortions in reducing the elastic energy and stabilizing the dimerized phase. For this reason, we think a resonance between the different bond-configurations shown in Figure 1 is more probable above $T_S$, than a random site-to-site fluctuation in the orbital occupancy. Therefore, resonating bonds are formed with the aid of unquenched orbital degrees of freedom, to remove the orbital degeneracy. At low temperature the structural distortion introduces an orbital anisotropy that is responsible for the static structure observed. Hence, a valence bond crystal state with fixed position of dimers is more appropriate for the low temperature phase, than a resonating valence bond configuration.

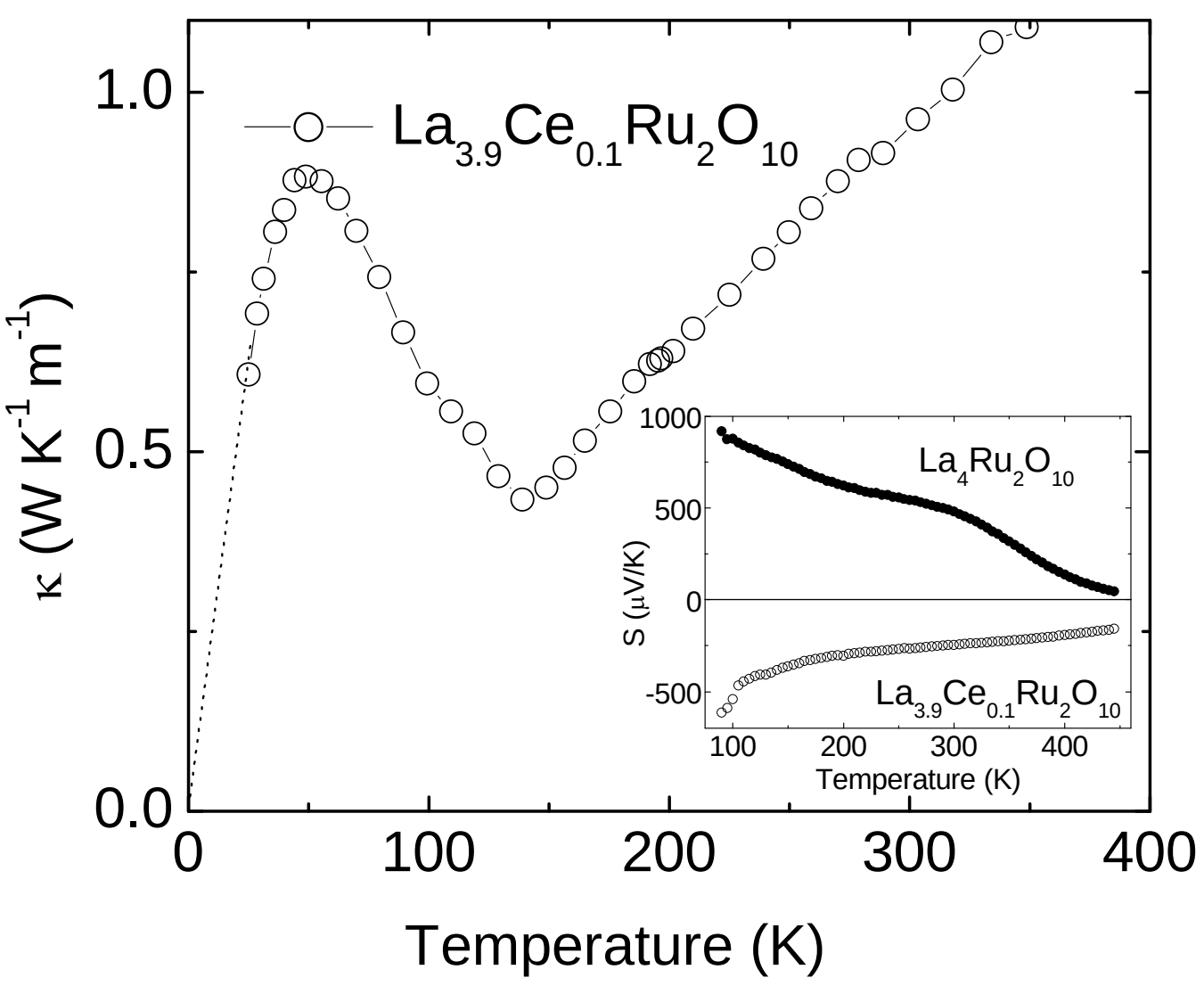

**Figure 5.** Temperature dependence of the thermal conductivity in $La_{3.9}Ce_{0.1}Ru_2O_{10}$. Inset: Thermoelectric power for the pure and Ce-doped sample. The hole-to-electron conduction after Ce-doping demonstrates the effective electron doping.

In this situation, the scenario predicted by equation (1) is completely modified above $T_S$: the Umklapp process must be largely suppressed, and the assumptions under Rayleigh dispersion mechanism are no longer applicable. In fact, when the defect is comparable in size or larger than the phonon wavelength, Joshi[18] obtained an inverse relaxation time which is dependent on $\omega^2$ instead of $\omega^4$. Making use of this frequency dependence, we obtained the linear κ(T), shown in Figure 3 (n=2, C = $4.5\times10^8$ s, A = $3.6\times10^9$ $s^{-1}$, B=0, in equation (2)), that fits the experimental data nicely above $T_S$.

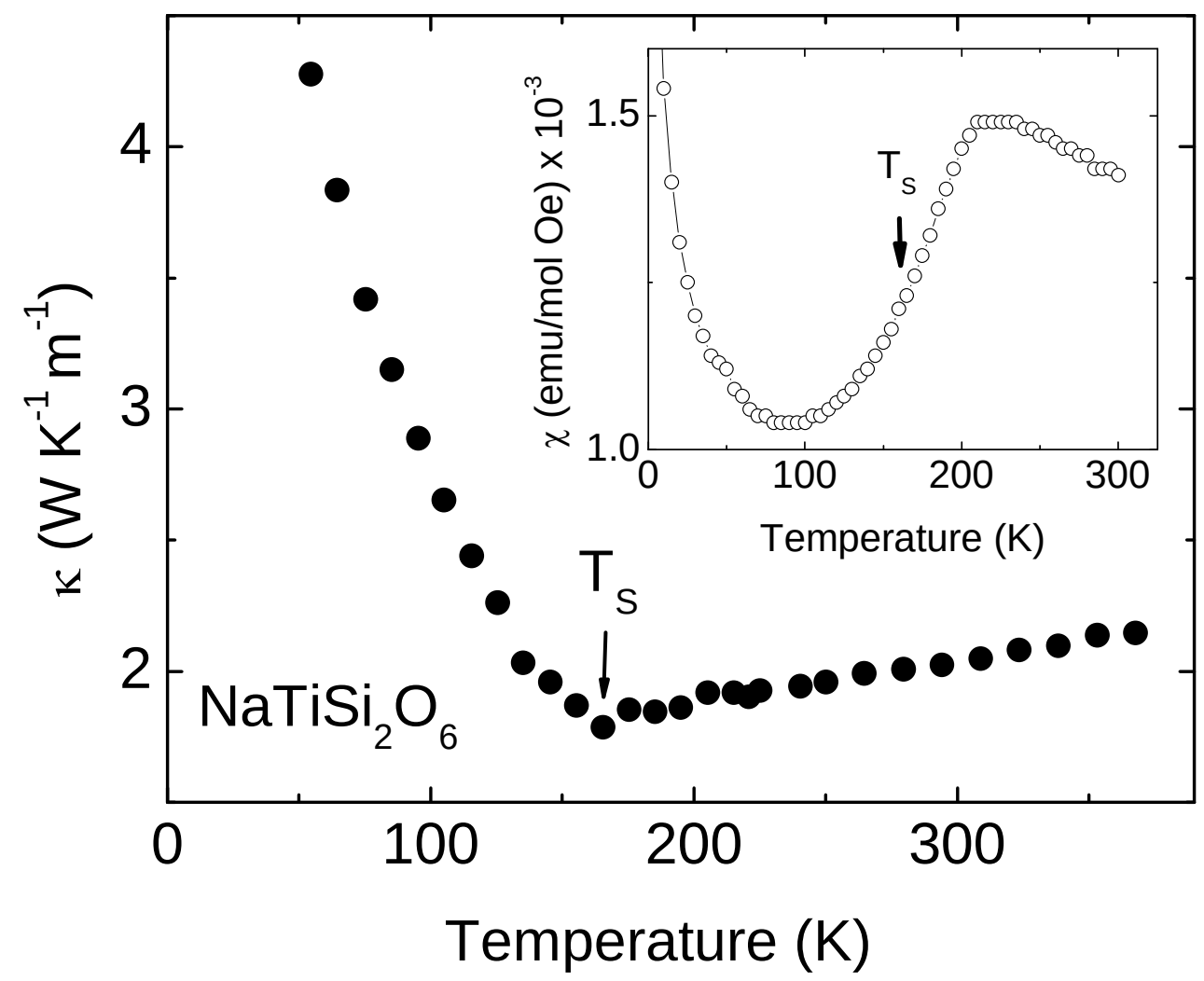

**Figure 6.** Temperature dependence of thermal conductivity for $NaTiSi_2O_6$. The vertical arrow marks the transition to the low temperature spin-singlet phase. Due to the large thermal conductivity at low

temperatures, we were unable to extend the measurements down to lower temperatures in our experimental setup. Inset: Temperature dependence of the magnetization at H=5T. The transition temperature, $T_s$, is defined from the maximum in the derivative of the M(T) curve.

Another source of linear $\kappa(T>T_s)$ in crystalline materials is the existence of Raman scattering, due to optical-acoustic coupling.[12,19] Although optical modes are not effective in transporting heat due to their small group velocity, they can interact with acoustic phonons under certain circumstances, influencing very much the heat transport. Dynamic disorder in systems with loosely bond guest species (clathrate, skuterrudites) reduces the energy of high-frequency lying optical modes and flattens the phonon bands, reducing the velocity of the acoustic phonons and therefore the thermal conductivity.[20] This mechanism introduces a dominant term in $\kappa \propto T$, which is qualitatively similar to the behavior of glasses and highly disordered systems like clathrate hydrates[21] and polymers,[19] and reproduces the temperature dependence we observed in the HT phase in $La_4Ru_2O_{10}$.

The distribution of phonons tends to be concentrated around a particular energy at a given temperature, which can be calculated $\approx 1.6k_BT$.[13] For a $\Theta \approx 225$ K, this corresponds to maximum phonon energy of $\approx$30 meV. Wu *et al.*[22] observed drastic changes in the optical phonon modes between $\approx$12 to 75 meV, showing that this optical-acoustic interaction is indeed a realistic possibility. At the light of our data, a complete description of the phonon spectrum of $La_4Ru_2O_{10}$ is highly desirable, for a quantitative fitting of $\kappa$ $(T>T_s)$.

Finally, in order to test the generality of this effect in other systems with orbital-induced spin-pairing, we have measured the thermal conductivity in $NaTiSi_2O_6$,[23] as shown in Figure 6. Above $T_s$, $\kappa(T)$ shows the linear behavior characteristic of glasses compatible with the persistence of dynamic bond-length distortions due to fluctuating orbital occupancy. The observation of an anomalous broadening of several phonon modes of the Raman spectra of $NaTiSi_2O_6$ above $T_S$,[24] can be better understood at the light of the present data.

Similar glass-like dependent thermal conductivity was reported in the orbitally disordered state of correlated perovskites ($AMnO_3$, $RVO_3$, etc), where semicovalent exchange striction introduces local bond-length fluctuations.[25]

Following these ideas, reducing even further the Jahn-Teller stabilization energy playing with $t_g$ electrons of 4d ions, combined with S or Se anions, could be a good place to look for systems in which the orbital liquid state is kept down to zero kelvin.[26]

Summarizing, the HT phase of $La_4Ru_2O_{10}$ is best described as a resonating mixture of different orbital orderings. This implies that orbital degrees of freedom are not quenched above the orbital-ordering transition, but remain in a state of rapid fluctuating occupancy. The same behavior is reported in the spin-gap compound $NaTiSi_2O_6$. Given the strong coupling among the orbital, lattice, spin and charge degrees of freedom that characterizes strongly correlated electron systems, liquid-like states in the spin and charge sectors are probable as well. Furthermore, this effect could be exploited to reduce the lattice thermal conductivity in correlated oxides, to foster their applicability in thermoelectric devices.

**Acknowledgments.** We acknowledge discussion with Prof. D. Khomskii, Dr. N. Perkins, Dr. J. Deisenhoffer, Dr. V. Pardo and Dr. I. Gonzalez. This work was supported by the Xunta de Galicia (INCITE09-209-101-PR). Facilities used by H.-D. Z. are supported in part by NSF through cooperative agreement No. DMR-0654118 and the State of Florida, USA.

---